\documentclass[preprint,review,12pt,authoryear]{elsarticle}

\usepackage{amssymb, amsmath}

\journal{}

\begin{document}

\begin{frontmatter}



\title{When does the stock market listen to economic news? New evidence from copulas and news wires.}


\author{Ivan Medovikov}

\address{Department of Economics, Brock University \\ 429 Plaza Building, 500 Glenridge Avenue, St. Catharines, Ontario, L2S 3A1, Canada. \\ Tel.: +1 905 688 55 50 ext. 6148, Email: imedovikov@brocku.ca}

\begin{abstract}
We study association between macroeconomic news and stock market returns using the statistical theory of copulas, and a new comprehensive measure of news based on the indexing of news wires. We find the impact of economic news on equity returns to be nonlinear and asymmetric. In particular, controlling for economic conditions and surprises associated with releases of economic data, we find that the market reacts strongly and negatively to the most unfavourable macroeconomic news, but appears to largely discount the good news. This relationship persists throughout the different stages of the business cycle.
\end{abstract}


\end{frontmatter}


\section{Introduction}
\label{section::introduction}

The impact of macroeconomic news on the stock market has been the subject of considerable amount of research during the past thirty years. Asset pricing theory suggests that news about macroeconomic factors such as employment or the level of output should influence financial markets, since it carries information about the aggregate investment opportunities set of the economy (see \cite{Merton1973} and \cite{Breeden1979}). Despite this, empirical evidence supporting market effects of real economic news remains weak and surprisingly mixed \footnote{Previous studies found statistically-significant association between stock market returns and various monetary variables such different interest rates (e.g. \cite{chen1986economic}, \cite{chen1991financial}, \cite{chan1998risk}) and monetary aggregates (e.g. \cite{cornell1983money}, \cite{pearce1983reaction}, \cite{pearce}). }. In many cases, important macroeconomic news is found to have no effect at all.  For example, \cite{flannery} examine the market impact of $17$ macroeconomic announcements and find that news relating to industrial production, unemployment and the real GNP appears to have no significant impact on stock prices. Similarly, earlier studies such as, for example, \cite{pearce} and \cite{Jain1988} find that the markets largely discount statistical announcements about industrial production and unemployment, even after taking prior expectations into the account. \cite{Ghent2010} also fails to find a significant market reaction to GDP and unemployment news. 

The different stages of the business cycle can change the context within which macroeconomic signals are interpreted, and the market impact of news becomes more apparent once the current economic conditions are taken into the account. \cite{Boyd2005} and \cite{mcqueen} find that news about rising unemployment has a negative effect on stock prices during economic contractions, since it can indicate lower corporate earnings and dividends, but a positive effect during expansions, as it may signal a greater likelihood of lower interest rates. More recently, \cite{Birz2011} find that after controlling for both the market expectations and the stages of the business cycle, the market reacts significantly to GDP and unemployment announcements. 
%

While the economic conditions associated with the market's interpretation and reaction to news are becoming better understood, interestingly, no studies appear to focus on news characteristics associated with greatest market impact. For example, controlling for economic conditions and expectations, does unusually-good or unusually-bad news lead to stronger market response? If so, is there asymmetry in this relationship? In other words, is the market effect of economic news non-linear and asymmetric? The aim of this paper is to fill this gap.

We re-asses the market effect of macroeconomic news using a relatively-new statistical methodology based on the theory of conditional copulas, and a new comprehensive measure of macroeconomic news that is derived from the indexing of news wires. Our measure quantifies polarity, intensity and volume of news related to U.S. employment, industrial activity, housing and construction, and the energy market. In addition to the news that relate to scheduled releases of economic statistics, our measure captures qualitative signals such as, for example, comments made by senior U.S. policy officials, policy developments, as well as man-made and natural disasters that may have an economic impact, and is arguably the broadest measure of economic news used in the literature to date. 

By adopting the copula approach, we are able to construct a flexible model that allows non-linear and asymmetric market effects of economic news. To our knowledge, this represents the first application of copulas to the analysis of the news-stock market relationship. Using the copula model, we map the association between economic news and the stock market in high detail and find that, controlling for expectations, surprises associated with releases of economic data, and prevailing economic conditions, the market impact of news is heavily skewed toward the most unfavorable announcements, which tend to be associated with significant market declines in all stages of the business cycle. In other words, it appears that it is only when the news is all doom and gloom, the market really listens. Lastly, since we're able to isolate the data-revealing component of news, we show that the qualitative economic news that our index captures does have a significant market effect. This may have implications for policy makers seeking to minimize potential disruptions associated with their announcements and comments. 

This paper is organized as follows. Section \ref{section::news_variable} details the construction of our news variable, which we call the Macroeconomic News Index. The index is based on manual review and classification of news-wire releases, and we review the data sources, indexing methodology, and the resulting index series in this section.  Section \ref{section::methodology} introduces the copula approach and details our modeling methodology. Section \ref{section::empirical_results} presents empirical results relating to the market impact of economic news. Lastly, a discussion of the results is provided in Section \ref{section::discussion}.

\section{Our news variable: the macroeconomic news index}
\label{section::news_variable}
Recent evidence suggests that the stock market reaction to economic news is not limited to responses to scheduled statistical releases. For example, while many failed to find any significant relationship between GDP announcements and the stock market in the past, \cite{Birz2011} document substantial market reaction to output-related news using a more comprehensive news measure based on the classification of newspaper headlines of \cite{Lott2004}.

We develop a new quantitative measure of U.S. macroeconomic news that is based on full review and classification of releases carried by major business news wires. In terms of economic coverage, professional news wire services differ from newspapers in several important ways, and we begin by detailing our news data sources and the indexing methodology used here.

\subsection{Sources of economic news}

We use the Thomson Reuters Newswires (TRN) and the Dow Jones Energy Service (DJES) as the sources of public information that underpins the news variable that we construct. Thomson Reuters is one of the largest news providers in the world, and maintains several newswires that cover a variety of U.S. business and economic developments. The Dow Jones Energy Service is a specialized source of news relating to the global energy industry, with coverage ranging from market information and significant firm-specific news to geopolitical events and energy policy. Since both are professional services, they emphasize timeliness of coverage and are usually among the first to break the story. Unlike the newspaper coverage, newswire releases tend to be highly-condensed and factual, making interpretation easier in most cases. They also contain few journalist's opinions, which substantially reduces political bias of economic coverage that has been found to exist in newspaper stories (e.g. see \cite{Mullainathan2005}, \cite{Groseclose2005}, \cite{dyck2003media} and \cite{Lott2004}). Their position at the top of the news chain, breadth of coverage and unambiguous format make the newswires an excellent source of information for indexing that is comprehensive, reduces the chance of misinterpretation, and contains little noise arising from the repetition of news by multiple outlets.

Since our aim is to capture and quantify the inflow of news relating to U.S. macroeconomic conditions, we restrict attention to releases that follow four key themes usually associated with economic activity. In particular, we focus on stories covering U.S. industrial production, national and regional labor markets, housing markets and construction industry, and the energy policy and markets. Industrial production, employment, and construction activity are often viewed as important indicators that carry signals about aggregate output, consumption, and return to capital. Energy costs can affect factor productivity and corporate profitability, which motivates our choice of energy news as one of index components.

We use the Dow Jones Factiva as the source of all historical newswire releases. Factiva captures and stores all messages carried by TRN and DJES in full, exactly as they appeared at the time of the release, without edits or revisions that may have been made to the content at a later date.  Updates and corrections to earlier releases are typically issued by TRN and DJES separately, and are also included into the index. 

The overall pool of stories transmitted during our sample period is extremely large, and only a minority of wires are related to U.S. macroeconomic conditions. We utilize the Dow Jones Intelligent Indexing (DJII) service to identify relevant releases that fit one of our four selected themes.

\subsection{Indexing methodology}

For every month in our sample, the construction of the index proceeds as follows. First, we collect all economic stories that were carried by TRN and DJES during the month and sort them into one of the four groups based on economic theme. For example, during April 2014 DJII search yields $27$ newswires relating to U.S. employment, $17$ wires covering U.S. industrial production, $7$ wires on housing markets and construction industry, and a further 27 releases focusing on energy markets and policy that were carried by TRN and DJES. Once the relevant stories are identified, we review the content of every story in full and classify the release either as a ``positive", a ``negative", or a ``neutral". A story is viewed as ``positive" if it contains information indicating improvement in current or future macroeconomic conditions, which for energy news is interpreted as easing of market conditions and a likely decline in price of key carriers such as oil or liquefied natural gas. ``neutral" stories are those that signal neither deterioration nor improvement. 

A large portion of industrial production, employment, and housing and construction-related newswires are driven by scheduled releases of macroeconomic statistics issued, among others, by U.S. Department of Commerce, the Bureau of Labor Statistics and Federal Reserve Banks, or by revisions to their previously-released figures. When interpreting such releases, it becomes important to account for prior expectations since, for example, a sound gain of $50,000$ in non-farm payrolls may in fact be viewed as negative news in times when higher growth was expected. Fortunately, in addition to the headline number, most data-related TRN releases also include a measure of expectations based on the most recent Reuters poll of forecasters. We therefore classify such releases in relation to expectations and treat the release as ``positive" if it exceeds expectations, ``negative" if it falls short of expectations, and ``neutral" if it meets the expectations. The part of the index that is driven by scheduled releases of core macroeconomic data therefore captures ``economic surprises", or the unexpected component of macroeconomic news.

Revisions to past economic data are another major component of data-driven news. For example, growth estimates issued by the U.S. Bureau of Economic analysis tend to be revised multiple times, and revisions can be substantial and occur several months after initial issue. Many historical macroeconomic data that are available today are therefore different form what was reported at the time of the release. The flow of such revisions represents important information that is absent from many contemporary data, and we include all news relating to revisions into the index.

The remaining, or non data-driven industrial production, employment, and housing and construction-related stories, along with the bulk of stories covering the energy markets, are qualitative in nature and include comments made by senior officials and policy-makers such as, for example, members of the board of governors of the Federal Reserve System, U.S. Secretary of Commerce or the Secretary of Energy, important geo-political developments such as sanctions or energy cartel quota decisions, as well as natural disasters with potential for economic impact. We describe the content of TRN and DJES economic releases along with specific classification guidelines in detail in the rest of this section. The total number of newswire releases reviewed during the construction of the index across the four news groups is presented in Table \ref{table::releasecountsummary}.
\begin{table}[htbp]
\begin{center}
\begin{tabular}{lrrrr}
\hline
\textit{News group} & \multicolumn{1}{l}{\textit{Positive}} & \multicolumn{1}{l}{\textit{Negative}} & \multicolumn{1}{l}{\textit{Neutral}} & \multicolumn{1}{l}{\textit{Total}} \\ \hline
\textit{Employment} & 3328 & 3220 & 479 & 7027 \\ 
\textit{Housing} & 982 & 830 & 126 & 1938 \\ 
\textit{Industry} & 1407 & 946 & 107 & 2460 \\ 
\textit{Energy} & 3830 & 3203 & 1281 & 8314 \\ \hline
\textit{Total} & 9547 & 8199 & 1993 & 19739 \\ \hline
\end{tabular}
\end{center}
\caption{Number of Reuters and Dow Jones Energy Service newswires by category, January 1999 to April 2014.}
\label{table::releasecountsummary}
\end{table}

Since our classification is similar to the grouping of responses to the University of Michigan Consumer Sentiment Index, we adopt its methodology and use the difference in percentage of positive and negative stories as the basis for the index. An analogous indexing approach is used in \cite{Birz2011}. To calculate the monthly value of the Macroeconomic News Index we first find the ratio of positive to negative stories in each of the four news groups. For example, Table \ref{table::subindexexample} shows the raw news counts and the resulting four \textit{sub-indexes} for April 2014. This yields a set of four indexes capturing polarity of U.S. employment, housing, industrial production, and energy-related news that are interesting in their own right, and are reviewed in more detail in Section \ref{section::macronewsindex}. The final value of the Macroeconomic News Index for the month is calculated by finding the average of the four sub-indexes, which amounts to $0.03$  during April 2014. It is evident that positive values of sub-indexes and of the final index indicate that a given month was dominated by positive news stories, while the magnitude of the reading is proportional to prevalence of certain type of news, with $1$ and $-1$ representing entirely positive and negative news-months respectively.
\begin{table}[htbp]
\begin{center}
\begin{tabular}{lrrrr}
\hline
 & \textit{Positive} & \textit{Negative} & \textit{Neutral} & \textit{Sub-index} \\ \hline
\textit{Employment} & 17 & 10 & 0 & 0.26 \\ 
\textit{Housing} & 0 & 5 & 2 & -0.71 \\ 
\textit{Industry} & 12 & 3 & 2 & 0.53 \\ 
\textit{Energy} & 13 & 13 & 1 & 0 \\ \hline
\end{tabular}
\end{center}
\caption{Raw news counts and resulting news sub-indexes, April 2014.}
\label{table::subindexexample}
\end{table}

The indexing methodology adopted here, as arguably other approaches to the building of such index, has some limitations. Here, the averaging of the four sub-indexes places equal weight on employment, housing, industry and energy-related news. Alternative weighting is clearly possible. To allow for alternative index definitions, we make our disaggregated data available in addition to the final index series. Next, we review the four sub-indexes and the resulting macroeconomic news index in more detail.

\subsection{Employment news sub-index}
\label{section::macronewsindex}
Nearly two thirds of employment-related wires cover scheduled releases and revisions to economic statistics by the agencies of the U.S. Department of Labor, with a large share of releases issued by the Bureau of Labor Statistics. These include statistics on national labor force, participation rates, employment and unemployment rates, non-farm payrolls, job openings and labor turnover, as well as jobless claims reports. State-level statistical releases also make a substantial part of the labor news flow and are included into the index, but their share relative to the national stories declines substantially closer to the end of our sample. Such data-driven releases contain least ambiguity, and interpretation is straightforward in most cases. For example, on August 1, 2014 Reuters wrote \begin{quote} ``U.S. job growth slowed a bit in July and the unemployment rate unexpectedly rose, pointing to slack in the labor market ... Nonfarm payrolls increased 209,000 last month after surging by 298,000 in June, the Labor Department said on Friday. Economists had expected a 233,000 job gain", \end{quote} which we recorded as ``negative" due to indication of raising unemployment and lower than expected job growth. 


Most non data-driven employment-related releases contain comments by senior officials on the state of the U.S. labor market. For example, on July 15, 2014, Reuters issued a release citing the Chair of the Federal Reserve: \begin{quote}
``Labor force participation appears weaker than one would expect based on the ageing of the population and the level of unemployment. These and other indications that significant slack remains in labor markets are corroborated by the continued slow pace of growth in most measures of hourly compensation",
\end{quote} which we also classify as ``negative". While such comments clearly indicate a negative view of the labor market, they may be perceived as good news in other context. For example, financial markets may view this as signal suggesting greater likelihood of monetary easing. Interaction between market impact of employment news and other economic variables is documented in, for example, \cite{Boyd2005}, where negative unemployment news is found to have a negative effect on equity prices during economic contractions, but a positive effect during expansions. Such non-linear effects can be captured using appropriate econometric models and controls, as we do in Section \ref{section::empirical_results}, rather than by contextual indexing of news. Here, we make particular effort to classify releases from the standpoint of underlying real economic signals, rather than its market interpretation or impact.

Figure \ref{fig::eni} of the Appendix shows the monthly values of employment news sub-index and the total number of employment-related news releases carried by Reuters between January 1999 and April 2014. While the index remained positive during most of the past decade and a half, the two periods of time dominated by persistent negative news coincide with $2001-2002$ and $2007-2009$ U.S. NBER recessions. The volume of labor market news varies substantially during our sample period reaching a peak of 83 releases in October 2002, with peak volume occurring roughly in the aftermath of both recessions, and on average amounts to 38 releases per month.

\subsection{Housing and construction news sub-index}
As with labor market news, most housing and construction releases are data-driven, and cover statistics by the U.S. Department of Commerce, National Association of Home Builders, National Association of Realtors, the American Institute of Architecture as well as the U.S. Department of Housing and Urban Development. Such releases typically include data on new and existing home sales, new home completions, building permits and housing starts, home financing costs, mortgage issue and refinance rates, home affordability and rents, and even architectural billings. Some data, such as home sales, building permits and completions are periodically revised, and we include news of revisions into the index. A substantial number of releases cover related commodity-market news such as, for example, significant changes in the price of construction timber. The remaining, non-data driven releases include news about major policy initiatives aimed at stimulating the construction industry and occasional extreme weather events that adversely affect home building.

\begin{figure}[ht]
\includegraphics[scale=0.8]{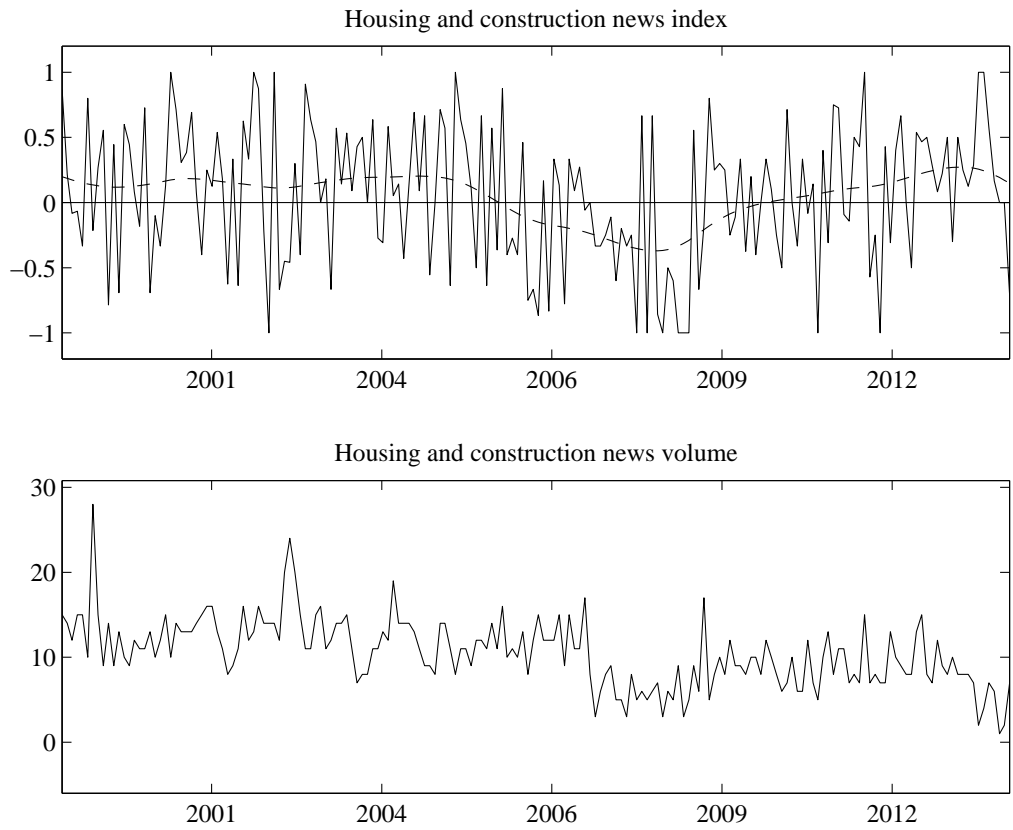}
\caption{Monthly values of housing and construction news index and total number of housing-related news releases carried by Reuters, January 1999 to April 2014. Dashed line shows HP-filtered trend.}
\label{fig::housingnewssubindex}
\end{figure}

The monthly values of the housing and construction news sub-index and the volume of housing-related news releases that appear in our sample are shown in Figure \ref{fig::housingnewssubindex}. As with the labor market news, the $1999 - 2014$ period was dominated by generally positive stories, with the exception of the five year span surrounding the U.S. housing crisis. The number of housing and construction-related releases carried by Reuters declines steadily and on average amounts to only $10$ releases per month, which is the lowest among the four news groups in the sample.

\subsection{Industry news sub-index}
The majority of signals about the U.S. industrial activity that receive news coverage are in the form of commentary, special reports, and statistical releases issued by the Federal Reserve banks, mainly of Chicago, Philadelphia and Richmond, industry associations such as the Institute of Supply Management (ISM) and the Manufacturers Alliance, as well as research circulated by major financial firms. These include data on industrial output, durable goods orders and non-farm productivity, along with a range of indexes of national and state-level manufacturing activity such as, for example, the Chicago Fed  Midwest Manufacturing Index or the PMI Purchasing Managers' Index. Periodic polls of analysts expectations by Reuters as well as data revisions also receive significant attention and are frequently mentioned in the news. As with employment and construction-related news, non-data driven industry releases mostly contain commentary by senior officials on the state of U.S. manufacturing and other industrial indicators.

\begin{figure}
\includegraphics[scale=0.8]{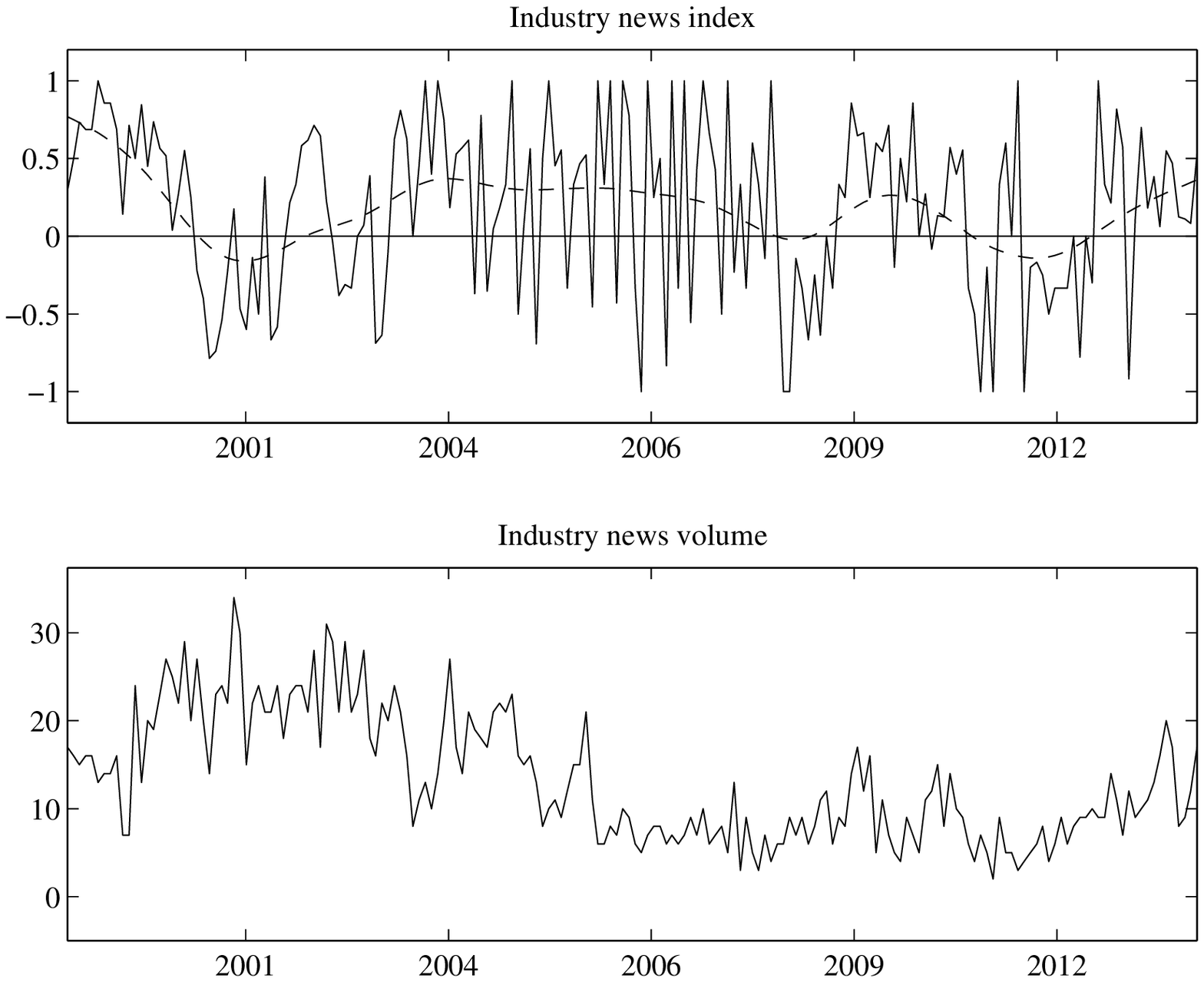}
\caption{Monthly values of the industry news index and total number of industry-related news releases carried by Reuters, January 1999 to April 2014. Dashed line shows HP-filtered trend.}
\label{fig::ini}
\end{figure}

The number of industry-related Reuters releases and the monthly values of the industry news sub-index are shown in Figure \ref{fig::ini}. As with employment news, periods or persistently-negative industry news appear to roughly coincide with the two most recent NBER recessions. Interestingly, the most recent span of negative industry news occurred in late $2011$ and early $2012$ -- well past the June $2009$ NBER recession end date, which supports the ``double-dip" view of the latest U.S. recession.

\subsection{Energy news sub-index}
Among the four groups of news included into the index, energy news contains the greatest number and by far the widest range of stories. Data-driven releases represent a minority of energy news and, in addition to announcements of major price changes for energy carriers such as crude oil or liquefied natural gas, include production figures, quota assignment and compliance rates by members of the OPEC, statistics on U.S. Strategic Petroleum Reserves, petroleum balances and other items tracked by the U.S. Energy Information Administration, as well as data released by the International Energy Agency. Non-data component of energy news is also very broad and contains news of natural and man-made disasters that adversely affect energy supply such as, for example, a major hurricane in the Gulf of Mexico, oil spill, or pipeline breakdown that interrupts oil production and refining, international and regional conflicts, as well as policy developments that lead to supply restrictions or easing such as sanctions or rules that permit new extraction.

\begin{figure}
\includegraphics[scale=0.8]{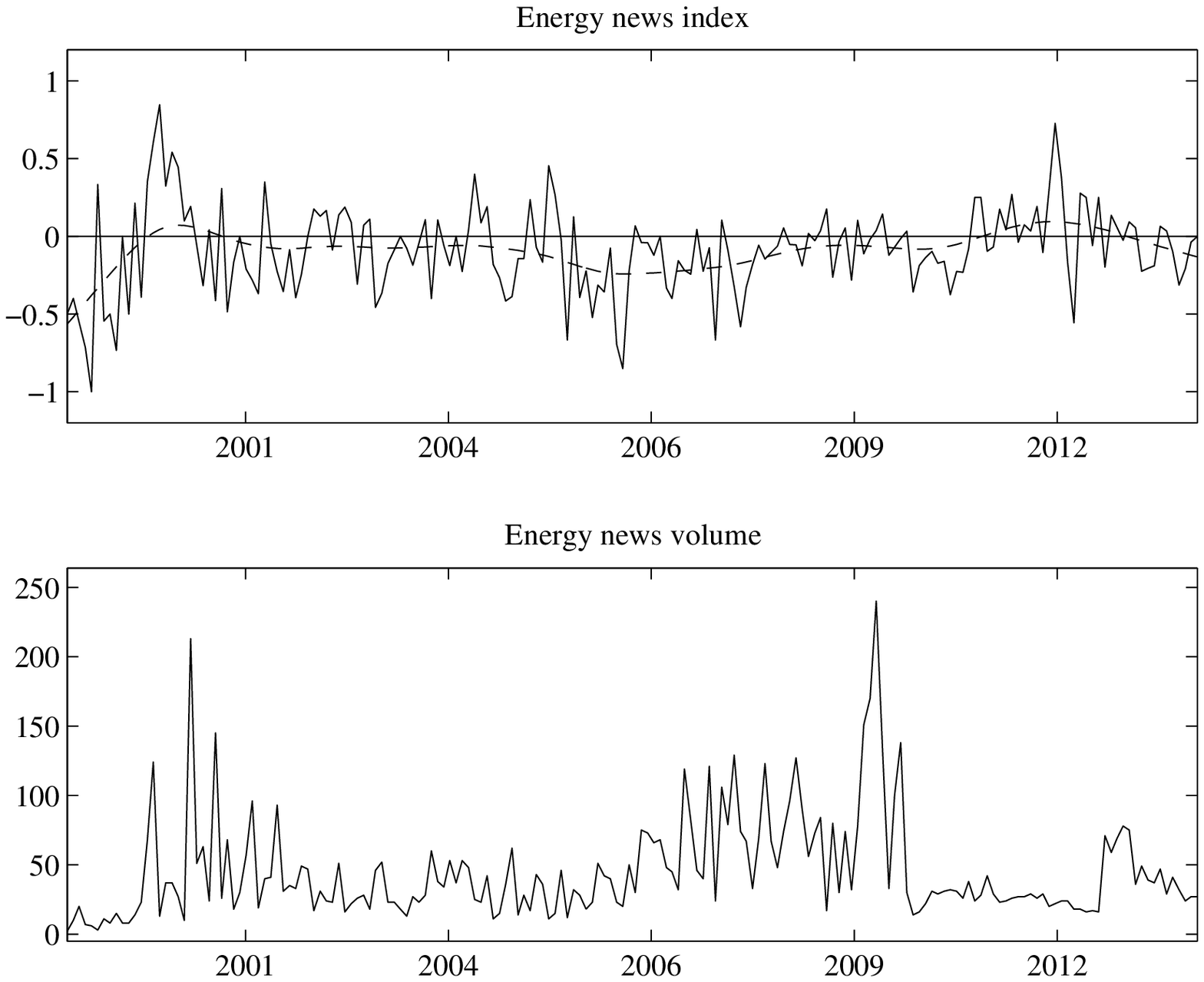}
\caption{Monthly values of the energy news index and total number of energy-related news releases carried by Reuters, January 1999 to April 2014. Dashed line shows HP-filtered trend.}
\label{fig::enni}
\end{figure}

Figure \ref{fig::enni} shows the number of energy-related DJES releases along with the values of the energy news sub-index. Set against the backdrop of steadily raising energy prices, combined with political turmoil and military action that involved some of the largest energy producers in the world, energy news sub-index remained negative throughout most of the past fifteen years, reflecting both growing supply risks and global energy demand.

\subsection{Combined index}
\begin{figure}
\includegraphics[scale=0.8]{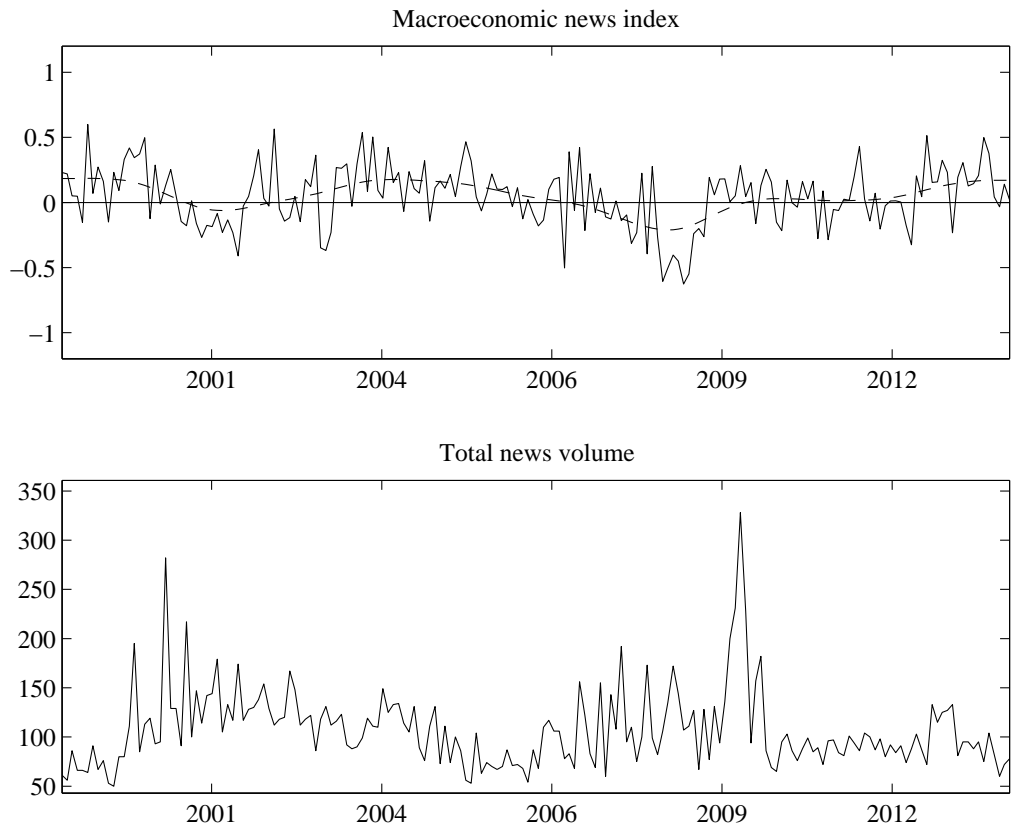}
\caption{Monthly values of the Macroeconomic News Index and total number of treated news releases carried by Reuters and Dow Jones Energy Service during January 1999-April 2014. Dashed line shows HP-filtered trend.}
\label{fig::mni}
\end{figure}

Final values of the Macroeconomic News Index, as well as the combined number of processed Reuters and DJIA releases, are shown in Figure~\ref{fig::mni}. The index appears to correctly reflect the inflow of poor economic news during the two NBER recessions during 1999-2014 and the prevalence of good news indicating periods of growth in between. 

\section{Our methodology: the conditional copula approach}
\label{section::methodology}
\subsection{A copula approach}

To gain a deeper understanding of the relationship between macroeconomic news and equity returns, we adopt a relatively-new statistical approach that is based on the theory of copulas. Consider a pair of random variables $X$ and $Y$, and let $F(x;\theta_x)$ and $G(y;\theta_y)$ represent their marginal distribution functions with parameters $\theta_x$ and $\theta_y$, and $H(x,y)$ be the joint CDF. Following a result by \cite{sklar}, the joint CDF can be expressed as
\begin{equation}
\label{eq::sklars}
H(x,y) = C[F(x;\theta_x),G(y;\theta_y);\theta_c],\hspace{5mm} (x,y) \in \mathbb{R}^2,
\end{equation} where $C:[0,1]^2\rightarrow[0,1]$ is the so-called copula of $X$ and $Y$, and $\theta_c$ is a vector containing copula parameters. Letting $u = F(x;\theta_x)$ and $v = G(y;\theta_y)$, it is evident that the copula is simply the joint CDF of $(u,v)$ which we can write as $C(u,v;\theta_c) = H[F^{-1}(u;\theta_x),G^{-1}(v;\theta_y)]$. 

Copulas are becoming central to the analysis of dependence as they provide a complete description of the association between $X$ and $Y$ that is also unique in the case that the variables are continuous. Different families of copulas represent a variety of dependence structures, with parameters in $\theta_c$ measuring the strength of association. For example, some of the better-known copula families include the Gaussian copula that captures linear correlation, Gumbel and Clayton copulas that measure asymmetric dependence that may be stronger among larger or smaller values of the data, as well as $t$ and Symmetrized Joe-Clayton (SJC) copulas that allow for tail dependence, or dependence among data extremes. 

Many well-known measures of association can be represented in terms of the copula. For example, rank-correlation measures such as Kendall's $\tau$ and Spearman's $\rho$ can be expressed in terms of $C$ as 
\begin{equation}
\tau = 4 \int_{0}^1 \int_{0}^1 C(u,v) dC(u,v) - 1
\end{equation} and 
\begin{equation}
\rho=12 \int_{0}^1 \int_{0}^1 C(u,v) dudv -3.
\end{equation}

A major advantage of the copula approach is that it enables the separate modelling of the marginal behaviour of $X$ and $Y$ and of the dependence structure embedded in $C$, meaning that a rich model of association that is free from limitations imposed by marginals $F$ and $G$ can be specified. For an introduction to copulas see \cite{Joe1997} and \cite{Nelsen2006}, and \cite{Cherubini2004} and \cite{Patton2009} for applications of copulas in finance. When the marginal models include other control variables, this amounts to the \textit{conditional copula} approach of \cite{Patton2006}. In this case, the copula captures dependence after ``netting out" the effects of variables included into the marginals as controls.

Our aim here is to explore the nature of dependence between our macroeconomic news variable and aggregate equity returns by specifying and fitting an accurate copula model to our data. While much work has been done to assess the impact of economic news on equity returns, our particular interest is in probing for non-linearities in this relationship such as tail dependence, which refers to dependence among extremes, or the tendency of very-large (or small) values of one variable to be associated with very-large (or small) values of another. In other words, in addition to gaining a better understanding of the overall association between macroeconomic news and security returns using the copula approach, our goal is to also measure the market impact of extreme news events. Such extreme dependence is usually studied through the so-called upper- and lower-tail dependence coefficients denoted $\lambda_u$ and $\lambda_l$ respectively and defined as 
\begin{eqnarray}
\lambda_u &=& \lim_{u\rightarrow 1^{-1}} Pr[F(x)\geq u | G(y) \geq u] = \lim_{u \rightarrow 1^{-1}} \frac{1 - 2u + C(u,u)}{1-u}, \\
\lambda_l &=& \lim_{u\rightarrow 0^+} Pr[F(x)\leq u | G(y) \leq u] = \lim_{u \rightarrow 0^+} \frac{C(u,u)}{u}.
\end{eqnarray} Greater values of $\lambda_u$ ($\lambda_l$) indicate stronger tendency of large (small) extremes of the variables to co-occur. One of our objectives is therefore to obtain estimates of coefficients $\lambda_u$ and $\lambda_l$ between our economic news variable and security market returns.

\subsection{Estimation}

Differentiating equation (\ref{eq::sklars}) will reveal that the joint PDF of $X$ and $Y$ can  be represented as 
\begin{equation}
f(x,y;\theta_x,\theta_y,\theta_c) = f(x;\theta_x) g(y;\theta_y) c(u,v;\theta_c),
\end{equation} where $f(x;\theta_x)$ and  $g(y;\theta_y)$ are the marginal densities  and $c(u,v;\theta_c)$ is the so-called \textit{copula density} defined as
\begin{eqnarray}
c(u,v;\theta_c) = \frac{\partial^2 C(u,v;\theta_c)}{\partial u \partial v}.
\end{eqnarray} Parameter estimates $\hat{\theta}_x$, $\hat{\theta}_y$ and $\hat{\theta}_c$ can therefore be obtained by maximizing the corresponding log-likelihood function 
\begin{equation}
\log(f(x,y;\theta_x,\theta_y,\theta_c)) = \log(f(x;\theta_x)) + \log(g(y;\theta_y)) + \log(c(u,v;\theta_c)),
\end{equation} where $l_x = \log(f(x;\theta_x))$ and $l_y = \log(g(y;\theta_y))$ are the log-likelihoods for the marginal models and $l_c = \log(c(u,v;\theta_c))$ is the \textit{copula log-likelihood}. 

\cite{Joe1996} propose a two-step method where marginal models are first estimated independently using maximum likelihood so that to obtain the values for $u$ and $v$, and estimates of copula parameters are obtained second by maximizing $l_c$, given the marginal estimates. When the marginals are specified parametrically, this method is known as Inference Functions for the Margins (IFM). When the marginals are non-parametric, this procedure is known as Canonical Maximum Likelihood (CML). Consistency and asymptotic normality of the IFM estimator under a set of regularity conditions is shown in \cite{Joe1997}. The estimator is also known to be highly-efficient (for example, see \cite{Michelis2010}), and we use it as the main estimation method here along with delete-one jack-knife procedure for estimating coefficient variances, as in \cite{Joe1996}.

\section{Empirical results}
\label{section::empirical_results}

We use the monthly dividend-inclusive return to a value-weighted portfolio of NYSE, NASDAQ and NYSE Arca-listed securities as our measure of aggregate U.S. equity returns, with all data taken from the Center for Research in Security Prices (CRSP) monthly stock files. Next, we specify and estimate the marginal models for security returns and the macroeconomic news index, and the copula model of dependence.

\subsection{Series diagnostic tests}

Our estimation approach requires that all series be stationary. In addition, to ensure validity of copula estimates, the marginal models must accurately capture any serial dependence such as autocorrelation or conditional heteroskedasticity that may be present in the series. We therefore begin our analysis by conducting diagnostic tests of our news variable $MNI_t$, value-weighted market returns $R_t$, Employment News Sub-Index $ENI_t$, Industry News Sub-Index $INI_t$, Energy News Sub-Index $ENNI_t$ and Housing and Construction News Sub-Index series $HNI_t$. Table \ref{table::diagnostictests} shows rejection decisions for the null-hypotheses of unit root based on the ADF test, conditional heteroskedasticity obtained with the ARCH test of \cite{Engle1982}, and serial correlation based on the Ljung-Box Q-test, for $12$ monthly lags of the series, all carried out at $5\%$ significance level. 

\begin{table}[htbp]
\begin{center}
\begin{tabular}{lccc}
\hline
 & \multicolumn{ 3}{c}{Reject null at $5\%$} \\ \hline
 & Unit Root  & Heteroskedasticity & Autocorrelation \\ \hline
$R_t$ & Reject & Do not reject & Reject \\ 
$MNI_t$ & Reject & Do not reject & Do not reject \\ 
$ENI_t$ & Reject & Do not reject & Do not reject \\ 
$INI_t$ & Reject & Reject & Do not reject \\ 
$ENNI_t$ & Reject & Do not reject & Do not reject \\ 
$HNI_t$ & Reject & Do not reject & Do not reject \\ \hline
\end{tabular}
\end{center}
\caption{Diagnostic tests for news varaibles and security returns.}
\label{table::diagnostictests}
\end{table}

We find that while none of the series appear to have a unit root, with the exception of $INI_t$, conditional heteroskedasticity is present in all series, and with the exception of $R_t$, all series are serially-correlated. 

\subsection{The marginal models}
\label{section::marginalmodels}
We begin with the marginal model for security returns. Similar to \cite{Birz2011}, we include an economic surprise variable into the marginal model for market returns to control for the effect of surprises associated with releases of economic statistics. Since the impact of news can depend on the current economic conditions, we also include the current unemployment rate to control for such interactions. As the returns are not serially-correlated, we do not include any lags. Our marginal model for security returns is then specified as follows:
\begin{eqnarray}
\label{eq::returnsmodel}
R_t &=& \beta_0 + \beta_1 X_t + \beta_2 L_t + \epsilon_{r,t}, \\
\sigma^2_{r,t} &=& \theta_0 + \theta_1 \sigma^2_{r,t-1} + \theta_2 \epsilon^2_{r,t-1}\text{, }\sqrt{\frac{1}{\sigma^2_{r,t}}}\epsilon_{r,t} \sim N(0,1),
\end{eqnarray} where $R_t$ is the value-weighted market return during the month $t$, $X_t$ is economic data surprises, $L_t$ is the current de-trended U.S. unemployment rate, and $\epsilon_{r,t}$ is the error term with variance $\sigma^2_{r,t}$, and $N(0,1)$ is the standard normal distribution. We therefore model $R_t$ as conditionally-normal, with time-varying mean that depends on the current economic conditions and real surprises, and variance that follows a $GARCH(1,1)$ process to capture volatility clusters. We use the U.S. economic surprise index of \cite{scotti2013surprise}, which shows accumulated differences between expected and actual economic data releases, as our measure of $X_t$. Controls for economic data surprises and current economic conditions here imply that our analysis of association between market returns and our economic news variable will be net of market reactions to surprises in statistical releases, and will also not be driven by interaction between the markets and the different stages of the business cycle. In other words, the dependence between equity markets and economic news that we assess in the next section is not driven by the data-revealing component of economic news, and is not due to the market effects of real economic fluctuations, but represents short-term market reactions to the various news-wire signals that we index.

Maximum likelihood estimates of the marginal model in (\ref{eq::returnsmodel}) are shown in Table \ref{table::returnmodelestimates}. 
\begin{table}[htbp]
\begin{center}
\begin{tabular}{lccccccc} \hline
 & $\hat{\beta}_0$ & $\hat{\beta}_1$ & \multicolumn{1}{c}{$\hat{\beta}_2$} & $\hat{\theta}_0$ & $\hat{\theta}_1$ & $\hat{\theta}_2$ & $R^2$ \\ \hline
\textit{Estimate} & 0.008 & 0.006 & 0.004 & 0.001 & 0.760 & 0.224 & 0.021 \\ 
\textit{t-Ratio} & \textbf{(2.65)} & (1.04) & (1.21) & (0.85) & \textbf{(8.92)} & \textbf{(2.56)} &  - \\ \hline
\end{tabular}
\end{center}
\caption{Maximum likelihood estimates of returns marginal model. Bold indicates significance at 5 s.l.}
\label{table::returnmodelestimates}
\end{table} As expected, the $GARCH(1,1)$ terms are highly-significant, confirming the presence of conditional heteroskedasticity in the return series. Much in line with previous literature, economic data surprises and unemployment are borderline-significant but still improve the overall model fit. 

The marginal model for our economic news variable is specified as follows:
\begin{eqnarray}
\label{eq::mnimodel}
MNI_t &=& \alpha_0 + \alpha_1 MNI_{t-1} + \alpha_2 MNI_{t-2} + \epsilon_{m,t},\\
\sigma^2_{m,t} &=& \gamma_0 + \gamma_1 \sigma^2_{m,t-1} + \gamma_2 \epsilon^2_{m,t-1}\text{, }\sqrt{\frac{1}{\sigma^2_{m,t}}}\epsilon_{m,t} \sim N(0,1),
\end{eqnarray} where $MNI_t$ is the value of the macroeconomic news index for month $t$ and $\epsilon_{m,t}$ is the error term that follows $GARCH(1,1)$ process as before. The number of lagged months of the news variable included into the model was determined first by estimating autoregressive model of order $6$ and then eliminating insignificant lags. Parameter estimates for model (\ref{eq::mnimodel}) are collected in Table \ref{table::mnimarginalmodelestimates}.

\begin{table}[htbp]
\begin{center}
\begin{tabular}{lccccccc}
\hline
 & $\hat{\alpha}_0$ & $\hat{\alpha}_1$ & $\hat{\alpha}_2$ & $\hat{\gamma}_0$ & $\hat{\gamma}_1$ & $\hat{\gamma}_2$ & $R^2$ \\ \hline
\textit{Estimate} & 0.025 & 0.243 & 0.279 & 0.019 & 0.497 & 0.088 & 0.19 \\ 
\textit{t-Ratio} & (1.51) & \textbf{(3.31)} & \textbf{(3.56)} & (0.69) & (0.73) & (0.698) & - \\ \hline
\end{tabular}
\end{center}
\caption{Maximum likelihood estimates of the marginal model for the macroeconomic news index. Bold indicates significance at $5\%$ s.l.}
\label{table::mnimarginalmodelestimates}
\end{table} Highly-significant estimates of auto-regressive coefficients indicate a substantial degree of persistence of economic news which is perhaps unsurprising given the persistent nature of business cycles.

\subsection{Goodness of fit tests for marginal models}
Before specifying and estimating the copula model, it is important to ensure that the marginals are correctly-specified. To this end, we use the $K-S$, $ARCH$ and $LBQ$ tests to probe for normality, homoskedasticity and absence of serial correlation of the residuals from models (\ref{eq::returnsmodel}) and (\ref{eq::mnimodel}) and find that we cannot reject any of the nulls at $5\%$ significance level for both models, indicating a good fit. 

Note also that probability transforms $u = F(x;\theta_x)$ and $v = G(y;\theta_y)$ should be uniformly distributed on $[0,1]$. Following \cite{Patton2006}, we calculate the transforms of the series $R_t$ and $MNI_t$ as $\hat{u}_t = \Phi(\hat{\epsilon}_{r,t} / \hat{\sigma}_{r,t})$ and $\hat{v}_t = \Phi(\hat{\epsilon}_{m,t} / \hat{\sigma}_{m,t})$, where $\Phi()$ is the standard normal CDF and $\hat{\epsilon}_{r,t} / \hat{\sigma}_{r,t}$ and $\hat{\epsilon}_{m,t} / \hat{\sigma}_{m,t}$ are the standardized residuals from the marginal models. As an additional goodness of fit check, we test the transforms for uniformity using the K-S test, and find that we cannot reject the null in both cases, with p-values being close to one.

\subsection{Empirical copula table}
To gain an initial understanding of the nature of association between security returns and economic news in our data, we construct the so-called \textit{empirical copula table} (for other examples see \cite{knight2005diversification} and \cite{ning2010dependence}). The copula table gives an overview of the structure of dependence, and is often used as the first step of the copula model selection process. First, we arrange the probability transforms $\hat{u}_t$ and $\hat{v}_t$ of the return and news index series in ascending order and then sort them evenly into four bins. In each case, the first bin contains the bottom $25\%$ smallest observations, and the fourth bin which contains the top $25\%$ largest observations in the sample. We then construct a frequency table with four rows and four columns, where the $j$ row and $i$ column shows the total number of elements in $j$'th bin of the news index series and $i$'th bin of the security return series. For example, the number in cell $(4,4)$ of this table shows the total number of observation pairs in the sample that are in top $25\%$ for both variables. It should be evident that such table represents the joint frequency distribution of probability transforms $\hat{u}_t$ and $\hat{v}_t$, and since these are uniform, our frequency counts should also be evenly distributed among the table cells when the variables are independent. That is, if security returns are independent from economic news, conditional on our controls, we should expect to see $11$ observations in each of the $16$ bins, since our monthly series spanning January-1999 to December-2013 contain a total of $180$ observations. A number greater than $11$ indicates a tendency of our variables to ``cluster" together in a particular bin. For example, a number in cell $(4,4)$ of this table that is substantially larger than $11$ would indicate that largest $25\%$ values of the news index tend to be associated with largest $25\%$ of market returns, and so on. 

\begin{table}[htbp]
\begin{center}
\begin{tabular}{c|cccc}
\hline
\multicolumn{1}{c|}{\textit{Bin}} & 1 & 2 & 3 & 4 \\ \hline
1 & \textbf{16} & 6 & 14 & 12 \\ 
2 & 10 & 8 & 12 & 10 \\ 
3 & 14 & 9 & 14 & 11 \\ 
4 & 8 & 12 & 13 & 11 \\ \hline
\end{tabular}
\end{center}
\caption{Empirical copula for security returns and macroeconomic news index.}
\label{table::empirical_copula_table}
\end{table}

We present the empirical copula table for our news variable and security market returns in Table \ref{table::empirical_copula_table}, with the largest deviation from independence count highlighted in bold. Interestingly, greatest deviation from expected count occurs among the bottom $25\%$ of news-return pairs where observed count exceeds expected by almost $50\%$, indicating that dependence between returns and economic news appears to be heavily skewed toward the lower tail of the joint distribution. The count in bin $(4,4)$ showing clustering of observations in top $25\%$ for both variables equals to that expected under independence, suggesting that little association exists among largest returns and news index values. In other words, the empirical coupla table appears to suggest that unusually-bad macroeconomic news tends to lead to substantial market declines, while equally unusually-good news shows no market effects.

\subsection{Copula model selection, estimation, and main result}
Asymmetric tail and nonlinear dependence between financial series has attracted some recent attention. For example, \cite{Patton2006} and \cite{Michelis2010} use copulas to study asymmetries in exchange rate dependence, and \cite{Ning2009} use a copula model to probe for tail dependence among security returns and trading volume. Here, since our marginal models include controls, our modelling approach is similar to the  conditional copula models of \cite{Patton2006} and \cite{Michelis2010}.

While our initial results indicate dependence that is skewed toward the lower distribution tail, we begin formal model selection by fitting several bi-variate copulas with a variety of dependence structures to the news index and market returns using the IFM method and assess their fit. We find that among Gaussian, Gumbel, Clayton, t and SJC coupla families, the one-parameter Clayton copula yields the superior fit on the basis of Akaike Information Criterion (AIC) and the Bayesian Information Criterion (BIC). The bi-variate Clayton copula is defined as
\begin{equation}
\label{eqn::copulamodel}
C(u,v;\theta) = \left [ \max(u^{-\theta} + v^{-\theta} - 1,0) \right ]^{-1/\theta}\text{, }(u,v)\in [0,1]^2,
\end{equation} where $\theta \geq 0$ is the dependence parameter such that greater values of $\theta$ indicate stronger association between the variables, and the case of $\theta = 0$ corresponds to independence. The Clayton copula features lower, but not upper-tail dependence, with tail-dependence coefficients given by $\lambda_l = 2^{-1/\theta}$ and $\lambda_u = 0$. Significant positive values of $\theta$ therefore also indicate the presence of lower-tail dependence.  The superior fit of the Clayton copula here supports our findings from the empirical copula table.

\begin{table}[htbp]
\begin{center}
\begin{tabular}{lccccc}
\hline
 & \textit{Estimate} & \textit{t-ratio} & \textit{AIC} & \textit{BIC} & \textit{LogL} \\ \hline
$\hat{\theta}$ & \multicolumn{1}{c}{0.134} & \multicolumn{1}{c}{\textbf{(1.899)}} & \multicolumn{1}{c}{-2.504} & \multicolumn{1}{c}{0.667} & \multicolumn{1}{c}{2.264} \\ \hline
\end{tabular}
\end{center}
\caption{IFM estimation results of bi-variate conditional Clayton copula model. Bold indicates significance at $5\%$ s.l.}
\label{table::copularesult}
\end{table}

IFM estimation results of the conditional Clayton copula model using our macroeconomic news index, value-weighted market returns and the marginal models specified in Section \ref{section::marginalmodels} are shown in Table \ref{table::copularesult}. The estimate of the Clayton dependence parameter is significant at $5\%$ s. l. indicating that, controlling for economic data surprises and persistence of news, macroeconomic news has a significant effect on security returns, and that this effect is skewed toward the lower distribution tail. In other words, we find that markets react strongly and negatively to extremely-poor economic news, but show no similar tendency to respond to good news, and that this relationship is not driven by the data-revealing content of economic news releases.

\subsection{Robustness checks}

In this section, we perform robustness checks to our result in Table \ref{table::copularesult}. Firstly, since our Macroeconomic News Index is calculated at monthly frequency, journalists writing the news-wires can observe market returns within the month, which may affect the tone of their reporting and create potential for reverse causation. Fortunately, TRN and DJES news-wires tend to be condensed and factual and contain little in terms of journalist opinions or subjective context. Classification of news-wires is therefore hardly affected by the reporter's tone. The index, however, reflects comments made by policy officials, which represents greater potential for reverse causation since such comments may be induced by market events in the first place. In other words, officials may issue commentary in response to market movements, which then gets objectively reported by TRN or DJIS and reflected in the index. Following \cite{Birz2011}, we note that if news is indeed driven by market activity, we should see more news releases during the high-activity months. To test this, we regress the absolute values of market returns on de-trended volume of news measured by the total number of news-wire releases in a given month and find the coefficient on news volume to be insignificant, with p-value close to one. Market activity is therefore not associated with volume of news. As a further test, we also re-estimate the marginal model for market returns given in (\ref{eq::returnsmodel}), but include de-trended number of news-wires as an additional control. As before, we find the news volume to be insignificant. As a final check, we keep the de-trended news volume in (\ref{eq::returnsmodel}) and re-estimate the entire Clayton copula model with IFM. Unsurprisingly, our estimation results change very little, and the Clayton dependence parameter remains significant with the new t-ratio of $1.91$. We therefore conclude that reverse causation between market returns and economic news is highly unlikely.
\section{Discussion}
\label{section::discussion}

It is interesting to note that since the probability transforms $\hat{u}_t$ and $\hat{v}_t$ are serially uncorrelated, the asymmetric market effect of news that we document here is not driven by any particular period of time such as, for example, the two NBER recessions in our sample, but is persistent throughout the business cycle. This may have implications for policy makers that have the potential to issue financially-disruptive comments. From the stock market standpoint, upbeat talk seems cheap, while economic pessimism can be costly at all times, not only during crises.

The news variable that we construct here may have some broader applications. For example, \cite{Engle2008} develop a model for low-frequency volatility that is driven by macroeconomic causes. Our news index may simplify the estimation of such models since it provides quantitative estimate of broader economic news that is available at a higher frequency than many scheduled statistical releases. 



  \bibliographystyle{elsarticle-harv} 
  \bibliography{macronews}

\newpage
\section{Appendix}

\begin{figure}[ht]
\includegraphics[scale=0.8]{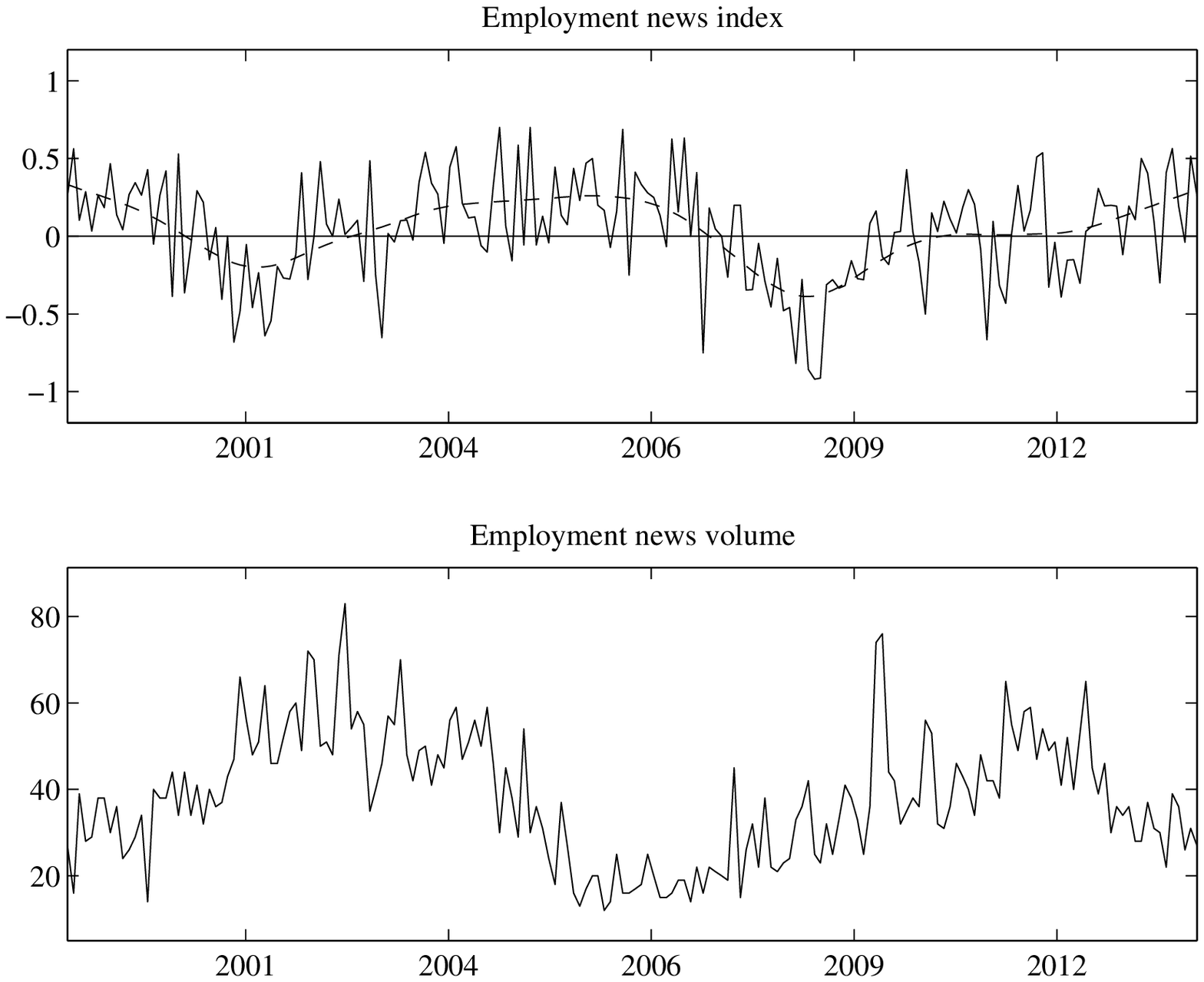}
\caption{Monthly values of employment news index and total number of employment-related news releases carried by Reuters, January 1999 to April 2014. Dashed line shows HP-filtered trend.}
\label{fig::eni}
\end{figure}

\end{document}